\documentclass[twocolumn,showpacs,preprintnumbers,amsmath,amssymb,aps,prd,nofootinbib]{revtex4}

\usepackage{graphicx}
\usepackage{bm} 

\begin{document}

\title{The $\bm r$-mode instability: Analytical solution with
gravitational radiation reaction}

\author{\'Oscar J. C. Dias}
\email{odias@perimeterinstitute.ca} \affiliation{Department of
Physics, University of Waterloo, Waterloo, Ontario N2L 3G1, Canada \\
                and \\
Perimeter Institute for Theoretical Physics, 31 Caroline St.\
N., \\
Waterloo, Ontario N2L 2Y5, Canada}
\author{Paulo M. S\'a}
\email{pmsa@ualg.pt} \affiliation{Departamento de F\'{\i}sica and
Centro Multidisciplinar de Astrof\'{\i}sica -- CENTRA, F.C.T.,
Universidade do Algarve, Campus de Gambelas, 8005-139 Faro,
Portugal}

\date{\today}

\begin{abstract}
Analytical $r$-mode solutions are investigated within the linearized theory
 in the case of a slowly rotating, Newtonian, barotropic,
non-magnetized, perfect-fluid star in which the gravitational
radiation (GR) reaction force is present. For the GR reaction term
we use the 3.5 post-Newtonian order expansion of the GR force, in
order to include the contribution of the current quadrupole
moment. We find the explicit expression for the $r$-mode velocity
perturbations and we conclude that they are sinusoidal with the
same frequency as the well-known GR force-free linear $r$-mode
solution, and that the GR force drives the $r$-modes unstable with
a growth timescale that agrees with the expression first found by
Lindblom, Owen and Morsink. We also show that the amplitude of
these velocity perturbations is corrected, relatively to the GR
force-free case, by a term of order $\Omega^6$, where $\Omega$ is
the angular velocity of the star.

\end{abstract}


\maketitle
\newpage
\section{Introduction}

Stars are not rigid systems; they naturally oscillate and the
non-radial pulsations generate gravitational radiation (GR) that
removes energy and angular momentum from the star. In non-rotating
stars this process is always dissipative and, inevitably, the
oscillation of the star dies off. However, if the star rotates
this is not necessarily the case: under certain conditions the
amplitude of the pulsation mode grows and a GR instability sets
in. In this paper we are interested on a special class of
GR-driven instabilities called Chandrasekhar-Friedman-Schutz (CFS)
instabilities, after the pioneering works of
Refs.~\cite{Chandra1970,FriedSchutz1978}. The generating mechanism
of these CFS instabilities is well understood. Indeed, consider a
mode that counter-rotates, i.e., whose propagation direction is
opposite to the star's rotation in the co-rotating frame of the
star. This mode has then negative angular momentum. When analyzing
the properties of the mode in the inertial frame, one sees that
the mode is dragged forward by the stellar rotation and can become
prograde, i.e., it can rotate in the same sense of the star. Then,
the gravitational waves emitted due to the mode oscillation carry
away positive angular momentum from the star. Since positive
angular momentum is extracted from a mode with negative angular
momentum, the amplitude of the mode increases and an instability
appears. So, although the GR emission lowers the inertial frame
energy, under the above conditions it induces an increase in the
mode energy as measured in the co-rotating frame. These two events
can occur simultaneously because the energy in the co-rotating
frame $E$ and the energy in the inertial frame $E_{\rm in}$ are
related by $E=E_{\rm in}-\Omega J$, where $\Omega$ is the angular
velocity of the star and $J$ is the angular momentum of the mode.
Therefore, $E$ can grow if both $E_{\rm in}$ and $\Omega J$
decrease \cite{FriedSchutz1978}. The condition for the appearance
of a CFS instability can be stated as follows. Given a mode with
an angular pattern speed $\sigma$ in the inertial frame that
propagates in a star that rotates with angular velocity $\Omega$,
a CFS instability sets in if $0<\sigma<\Omega$. In general, this
means that for a given mode with fixed $\sigma$ the instability
will appear only if the star is rapidly rotating with an angular
velocity greater than the critical value $\Omega_{\rm
crit}=\sigma$. This requirement of a large stellar rotation,
together with the unavoidable viscosity contributions that damp
the instability, has diminished the astrophysical interest in the
CFS instabilities \cite{Lind1995}.

This pessimistic scenario changed when the properties of a special
class of stellar pulsation modes, called $r$-modes, were
investigated in detail. The $r$-modes, which were first studied
\cite{PapaPringle1978,ProvostBerthRocca1981,Saio1982,SmeyMar1983}
more than twenty years ago, are pulsation modes of rotating stars
that have the Coriolis force as their restoring force. These modes
induce perturbations mainly in the fluid's velocity and cause very
small disturbances in the star's density. More precisely, the
velocity perturbations are of order $\Omega$, while the density
perturbations are of order $\Omega^2$, where  $\Omega$ is the
star's angular velocity. The unique feature that clearly
distinguishes the $r$-modes from other stellar modes and that turn
them potentially interesting for astrophysics is the fact that
they are driven unstable by GR reaction in all perfect-fluid
rotating stars, no matter how slowly they rotate. This property
was discovered by Andersson \cite{Andersson1998}, and afterward
confirmed more generally by Friedman and Morsink
\cite{FriedMors1998}. So, unlike other modes, $r$-modes are always
retrograde in the star's co-rotating frame and prograde in the
inertial frame, i.e., the sign of the $r$-mode frequencies is
always opposite in the two frames. In other words, the CFS
instability always occurs for the $r$-modes because the angular
pattern speed of a $r$-mode with angular quantum number $l$ in an
inertial frame is given by
$\sigma=\frac{(l-1)(l+2)}{l(l+1)}\,\Omega$ \cite{PapaPringle1978},
and so it satisfies the CFS condition, $0<\sigma<\Omega$, for any
value of the stellar rotation and for $l\geq 2$.

Two questions are then naturally raised, namely, what are the
timescales associated with the growth of the $r$-mode CFS
instability, and are the several viscosity processes and damping
mechanisms  present in a star sufficient to suppress the GR
instability? These questions were first addressed in
Ref.~\cite{LindOwenMors1998}, where, for the evaluation of the
growth timescale $\tau$ of the instability, the small density
perturbations induced by the $r$-mode were neglected and, in the
framework of Newtonian hydrodynamics, the formalism developed in
Ref.~\cite{Thorne1980} was used to compute the GR reaction that
induces the GR instability. GR couples to $r$-modes primarily
through the current multipoles rather than the usual mass
multipoles (this result follows straightforwardly from an order
count in the multipoles: the perturbations in the mass multipoles
are of order $\Omega^2$, while the perturbations in the current
multipoles are of order $\Omega$). In Ref.~\cite{LindOwenMors1998}
it was found that the energy $E$ of the $r$-mode perturbation in
the co-rotating frame varies, to lowest order in $\Omega$,
according to $E=E_0 e^{-2t/\tau}$, where $E_0$ is the initial
energy. Moreover, for $l=2$, the timescale $\tau_{\rm GR}$ of the
GR-driven instability is given by \cite{LindOwenMors1998}
\begin{eqnarray}
\tau_{\rm GR} = - \left(
    \frac{2^{17}\pi}{5^2\times3^8} \, \frac{G}{c^7} \, \tilde{J} \Omega^6
                          \right)^{-1}\,,
 \label{GRtimescale:LindOwenMors}
\end{eqnarray}
where $G$ is the Newton's constant, $c$ is the velocity of light,
and $\tilde{J}\equiv \int_0^R dr \rho r^6$ with $\rho$ and $R$
being the density and the surface's radius of the unperturbed
star, respectively. In parallel, a similar programme as the one of
Ref.~\cite{LindOwenMors1998} was carried in
Ref.~\cite{AndersKokkSchutz1999}, but with the important
difference that, in order to include the small density
perturbations, the calculation was performed up to second order in
the stellar rotation ($\Omega^2$). The density perturbation
contribution is important to correctly evaluate the bulk viscosity
effects, but has no influence on the GR timescale $\tau_{\rm GR}$
\cite{AndersKokkSchutz1999}. The calculations performed in
Refs.~\cite{LindOwenMors1998,AndersKokkSchutz1999} indicate that,
for a wide range of relevant temperatures and angular velocities
of the star, $\tau_{\rm GR}\ll \tau_{\rm V}$, with the
gravitational timescale $\tau_{\rm GR}$ being several orders of
magnitude smaller than the viscous timescale $\tau_{\rm V}$. Thus,
for this range of temperatures and angular velocities of the star,
viscosity cannot suppress the growth of the GR-driven instability
of $r$-modes.

The astrophysical implications of the GR-driven instability in the
$r$-modes are then very promising. Indeed, a long-standing
unsolved problem of pulsar astrophysics is associated with the
observed slow rotation rates of neutron stars. Neutron stars are
believed to be formed through stellar gravitational collapse. The
angular momentum of the system is conserved during this process
and so the natural expectation is that the final neutron star
should have a rotation rate close to the Keplerian velocity, i.e.,
the maximum velocity above which matter starts escaping through
the equatorial plane (see nevertheless
Ref.~\cite{SpruitPhinney1998} for a different view). However, all
the observed neutron stars present a rotation rate well below this
maximum value. The $r$-modes with their peculiar properties
provide a possible explanation for these small rotation rates of
young pulsars in supernova remnants. Indeed, as we just saw, in a
newly born, hot, rapidly-rotating neutron star the GR reaction
force dominates bulk and shear viscosities for enough time to
allow most of the star's angular momentum to be radiated away as
gravitational waves \cite{LindOwenMors1998,AndersKokkSchutz1999}.
As a result, the neutron star might spin down to just a small
fraction of its initial angular velocity. The $r$-modes may also
play an important astrophysical role in low-mass X-ray binaries.
Indeed, GR emission due to the $r$-mode instability could balance
the spin-up torque due to accretion of neutron stars in low-mass
X-ray binaries, limiting in this way the maximum angular velocity
of these pulsars to values consistent with observations
\cite{Bildstein1998,AndersKokkSterg1999,Chak2003}. GR released
during the $r$-mode evolution seems also to be a very promising
source for the detection of gravitational waves, even though the
perturbations in the density induced by $r$-modes are small and
thus, in a first intuitive analysis, they should not be associated
with strong GR emission. A first attempt to develop a
time-dependent model of the evolution of the $r$-mode instability,
including an approximate discussion on the nonlinear phase where
most of the angular momentum of the star is radiated away, has
shown that the GR emitted can probably be detected in enhanced
versions of laser interferometer detectors \cite{OwenAl1998}.
Although recent results on the nonlinear saturation of the
$r$-mode energy are not so optimistic, they still point to the
possible detection of gravitational waves if the unstable neutron
stars are close enough \cite{ArrasAl2003}.

A quite interesting feature that has emerged from recent
investigations on $r$-modes is the presence of differential
rotation induced by the $r$-mode oscillation in a background star
that is initially uniformly rotating. That differential rotation
drifts of kinematical nature could be induced by $r$-mode
oscillations of the stellar fluid was first suggested in
Refs.~\cite{RezzLambShap1,RezzLambShap2}, where an approximate
analytical evaluation of nonlinear effects was performed.
Differential rotation drifts induced by $r$-modes were also found
in numerical simulations of nonlinear $r$-modes carried out both
in general relativistic hydrodynamics \cite{StergFont2001}, and in
Newtonian hydrodynamics
\cite{LindTohVal2001,LindTohVal2002,GressmanEtAll2003}.
Differential rotation was also reported in a model of a thin
spherical shell of a rotating incompressible fluid
\cite{LevinUsh2001}. Recently, an analytical solution,
representing differential rotation of $r$-modes that produce large
scale drifts of fluid elements along stellar latitudes, was found
within the nonlinear Newtonian theory up to second order in the
mode amplitude and in the absence of GR reaction \cite{Sa2004}.
This work is exact to second order, improving the approximate
analytical computation performed in
Refs.~\cite{RezzLambShap1,RezzLambShap2}. As shown in
Ref.~\cite{SaTome2004}, this differential rotation plays a
relevant role in the nonlinear evolution of the $r$-mode
instability. Indeed, in a newly born, hot, non-magnetized, rapidly
rotating neutron star, $r$-modes saturates a few hundred seconds
after the mode instability sets in; the saturation amplitude
depends on the amount of differential rotation at the time the
instability becomes active and can take values much smaller than
unity \cite{SaTome2004}. Now, the question that remains unsolved
is whether GR reaction also induces differential rotation, making
an extra contribution. One of the aims of the present paper is to
initiate a programme that hopefully will allow to answer this
question.

In this paper, we will perform an analytical study of $r$-mode
solutions within the linearized theory  in the case of a slowly
rotating, Newtonian, barotropic, non-magnetized, perfect-fluid
star in which a gravitational radiation reaction force is present.
The GR reaction contribution is assumed to be given by the 3.5
post-Newtonian order expansion that includes the current
quadrupole moment, which is the main responsible for the GR
instability that sets in. The detailed formalism of the 3.5
post-Newtonian order expansion is given in
Refs.~\cite{Blanchet1997,Rezzolla1999}. It is known that the most
unstable mode is the $l=m=2$ $r$-mode, where $l$ and $m$ are the
angular momentum and the azimuthal numbers, respectively. Hence,
we will focus our attention on the GR-driven instability of this
mode. We will find an analytical expression for the $r$-mode
velocity perturbations, which is physically consistent with the
energy evolution studied in
Refs.~\cite{LindOwenMors1998,AndersKokkSchutz1999}.

The plan of the paper is the following. In
Section~\ref{FullTheory} we write the full Newtonian hydrodynamic
equations with the GR reaction force. We review the main
properties of the GR reaction force and we give the general
formulas needed to compute this reaction force. The explicit
computation of the GR reaction force in the 3.5 post-Newtonian
order expansion is then performed in Appendix~\ref{GRforce}.
Section~\ref{LinearTheory} is devoted to the linearized theory.
The perturbed Newtonian hydrodynamic equations with the GR
reaction force are written in Subsection~\ref{LinearHydroEqs}. In
Subsection~\ref{Linear r-mode noGRforce} we briefly review the
well-known GR force-free linear $r$-mode solution that is used in
Subsection~\ref{EulerVariationGRforce} to compute the first-order
Eulerian change in the GR force. In Subsection~\ref{Linear r-mode
GRforce} we finally solve the Newtonian hydrodynamic equations
with the GR force. In Section~\ref{Conclusion} our results are
discussed and future directions are pointed. In
Appendix~\ref{Comparing}, we identify explicitly the connection
between some of our intermediary formulas for the GR force and
those of previous numerical works on the subject.

\section{Newtonian hydrodynamic equations with
GR reaction force} \label{FullTheory}

The Newtonian hydrodynamic equations for a uniformly rotating,
barotropic, non-magnetized, perfect-fluid star in the presence of
the gravitational radiation (GR) reaction force are the Euler,
continuity, and Poisson equations given, respectively, by
\begin{eqnarray}
&
\partial_t \vec{v} + ( \vec{v} \cdot \vec{\nabla} ) \vec{v}
 = - \rho^{-1} \vec{\nabla} P - \vec{\nabla} \Phi + \vec{F}^{\rm GR},
&
 \label{Euler}
\\
&
\partial_t \rho + \vec{\nabla} \cdot (\rho \vec{v}) = 0,
&
 \label{Continuity}
\\
&
 \nabla^2 \Phi = 4 \pi G \rho,
&
 \label{Poisson}
\end{eqnarray}
where $\rho$ is the density of the fluid, $\vec{v}$ is its
velocity, $P$ is its pressure, $\Phi$ is the Newtonian potential,
and $\vec{F}^{\rm GR}$ is the GR reaction force per unit mass (or,
simply, GR reaction force). The general expression for the GR
reaction force $\vec{F}^{\rm GR}$, that includes contributions
from the time-varying mass multipole moments and from the
time-varying current multipole moments, has been found and
progressively discussed in detail up to the $3.5$ post-Newtonian
order. A review of this path and the $3.5$ post-Newtonian
expansion of the GR force can be found in
Refs.~\cite{Blanchet1997,Rezzolla1999}. In this paper we are
interested on the contribution of the GR reaction force to the
$r$-mode oscillations. Contrary to what happens in other mode
oscillations, the $r$-mode instability is predominantly excited by
the time-varying current multipole moments rather than by the
usual time-varying mass multipole moments \cite{LindOwenMors1998}
(see also a detailed discussion in Ref.~\cite{Rezzolla1999}). Now,
in order to include the contribution of the current multipole
moments, the post-Newtonian expansion of the GR reaction force
must be done at least up to the 3.5 order
\cite{Blanchet1997,Rezzolla1999}. This post-Newtonian order
includes the contribution of the current quadrupole moment, which
is the dominant current multipole moment. Since the main
contribution for the $r$-mode instability comes from the current
quadrupole moment, we will henceforth neglect all the terms coming
from the time-varying mass multipole moments, i.e., we will not
consider terms that appear in post-Newtonian orders below 3.5
neither the terms of the 3.5 post-Newtonian order that have their
origin in the time-varying mass multipole moments. These terms are
clearly identified in Refs.~\cite{Blanchet1997,Rezzolla1999}. So,
in the so-called Blanchet's gauge, the contribution of the current
multipole moments to the 3.5 post-Newtonian order  GR force,
henceforth labelled as $\vec{F}^{\rm GR}$ for simplicity, is
\footnote{See equations (12) and (17) of Ref.~\cite{Rezzolla1999}.
Our notation is slightly different from the one of
Refs.~\cite{Blanchet1997,Rezzolla1999}. The correspondence is
$\vec{\beta}\equiv _{8}\!\!\vec{\beta}$.}
\begin{equation}
 \vec{F}^{\rm GR} = - \partial_t \vec{\beta} + \vec{v} \times
 ( \vec{\nabla} \times \vec{\beta} ),
 \label{DefGravForce}
\end{equation}
where $\vec{\beta}$ is a vector whose components are given by
\begin{equation}
 \beta_i \equiv \frac{16G}{45c^7} \epsilon_{ijk} x_j x_q S_{k q}^{[5]},
 \label{DefBeta}
\end{equation}
with $i=1,2,3$ and similarly for the other Latin indices. In the previous
expressions, $S_{i j}(t)$ is the time-varying current quadrupole tensor,
\begin{equation}
 S_{i j}(t) \equiv \int d^3x \epsilon_{kq(i}x_{j)} x_k \rho v_q,
 \label{DefSij}
\end{equation}
$\epsilon_{i j k}$ is the Levi-Civita tensor, $x_i$ is the
Cartesian coordinate of the point at which the tensor is
evaluated, and
\begin{equation}
 S_{i j}^{[n]}(t) \equiv \frac{d^n}{dt^n} S_{i j}(t).
 \label{DefDerivative}
\end{equation}
The explicit computation of the GR reaction force is performed in
Appendix \ref{GRforce}.

It is interesting to note that $\vec{\beta}$ can be appropriately identified
as a gravitational vector potential \cite{LevinUsh2001}. This designation
comes from the clear equivalence between the GR force per unit mass exerted on
a moving fluid (\ref{DefGravForce}) and the Lorentz force per unit charge
exerted by an electromagnetic field on a moving charged fluid, $\vec{F}^{\rm
Lor}= -\partial_t \vec{A}+ \vec{v}\times (\vec{\nabla} \times \vec{A})$, where
$\vec{A}$ is the electromagnetic vector potential. In this analogy, the
designation of  gravitational vector potential for $\vec{\beta}$ follows
straightforwardly if one identifies $\vec{\beta}\equiv \vec{A}$.

\section{$\bm R$-modes in the linearized theory with the
GR reaction contribution} \label{LinearTheory}

In previous investigations
\cite{LindOwenMors1998,AndersKokkSchutz1999}, the hydrodynamics
equations with the GR force were used to obtain an expression for
the time evolution of the physical energy of the $r$-mode
perturbation, $dE/dt$, from which the gravitational radiation and
viscous timescales were determined. In this section, we explicitly
solve the hydrodynamics equations with the GR force, obtaining an
expression for the first-order Eulerian change in the velocity of
the $r$-modes $\delta^{(1)} \! \vec{v}$. It is shown that these
velocity perturbations are sinusoidal with the same frequency as
the well-known GR force-free linear $r$-mode solution; that the GR
force drives the $r$-modes unstable with a growth timescale that
agrees with the expression found in
Refs.~\cite{LindOwenMors1998,AndersKokkSchutz1999}; and that the
amplitude of these velocity perturbations is corrected, relatively
to the GR force-free case, by a term of order $\Omega^6$.

\subsection{Linearized hydrodynamic equations}
\label{LinearHydroEqs}

The hydrodynamic equations (\ref{Euler})--(\ref{Poisson}) for a
uniformly rotating, Newtonian, barotropic, non-magnetized,
perfect-fluid star with the GR reaction force can be linearized
yielding, in the inertial frame,
\begin{eqnarray}
& \hspace{-2 cm}
\partial_t \delta^{(1)} \! \vec{v} + ( \delta^{(1)} \! \vec{v} \cdot
\vec{\nabla} ) \hat{\vec{v}} + ( \hat{\vec{v}} \cdot \vec{\nabla}
) \delta^{(1)} \! \vec{v} \nonumber
 \label{EulerLinearized}
&
\\
& \hspace{3 cm}= - \vec{\nabla} \delta^{(1)} \! U + \delta^{(1)}
\! \vec{F}^{\rm GR}, &
\\
&
\partial_t \delta^{(1)} \! \rho + \hat{\vec{v}} \cdot \vec{\nabla}
\delta^{(1)} \! \rho + \vec{\nabla} \cdot \left( \hat{\rho}
\delta^{(1)} \! \vec{v} \right) = 0,
 \label{ContinuityLinearized}
&
\\
&
 \nabla^2 \delta^{(1)} \! \Phi = 4\pi G \delta^{(1)} \! \rho,
 \label{PoissonLinearized}
&
\end{eqnarray} where we have defined
\begin{equation}
\delta^{(1)} \! U \equiv \frac{\delta^{(1)} \! P}{\hat{\rho}} +
\delta^{(1)} \! \Phi. \label{DefU}
\end{equation}
In the above equations,
\begin{equation} \hat{\vec{v}}=\Omega r \sin\theta
\vec{e}_{\phi} \label{UnperturbVeloc}
\end{equation}
is the velocity of the unperturbed star and $\hat{\rho}$ its mass
density, while $\delta^{(1)} \! Q$ denotes the first-order
Eulerian change in a quantity $Q$.

\subsection{Linear $\bm r$-mode solution
without the GR reaction contribution} \label{Linear r-mode
noGRforce}

In our approach, we use the linear $r$-mode solution without the
GR contribution to generate the GR reaction force. As already
mentioned, the GR force drives $r$-modes unstable and the most
unstable mode is the $l=m=2$ $r$-mode. We will therefore
concentrate our attention in this mode and in this subsection we
briefly present this well-known solution.

In the absence of the GR reaction force, $\delta^{(1)} \!
\vec{F}^{\rm GR}=0$, and to lowest order in $\alpha$ and in
$\Omega$, where $\alpha$ is the dimensionless $r$-mode amplitude,
the perturbed equations
(\ref{EulerLinearized})--(\ref{PoissonLinearized}) allow $r$-mode
solutions with velocity perturbations given by $\delta^{(1)} \!
\vec{v} = \alpha \Omega R (r/R)^l \vec{Y}^B_{ll} e^{i\omega t}$,
where $R$ is radius of the unperturbed star. In spherical
coordinates and for $l=2$, the velocity perturbations are
explicitly given by
\begin{subequations}
 \label{r-mode:noGRforce}
\begin{eqnarray}
 \delta^{(1)} \! v_r &=& 0,
\\
 \delta^{(1)} \! v_{\theta} &=& -\frac{i}{4} \sqrt{\frac{5}{\pi}}
\frac{\alpha\Omega}{R} r^2
 \sin\theta e^{i(2\phi+\omega t)},
\\
 \delta^{(1)} \! v_{\phi} &=& \frac14 \sqrt{\frac{5}{\pi}} \frac{\alpha\Omega}{R}
   r^2 \sin\theta \cos\theta
 e^{i(2\phi+\omega t)},
\end{eqnarray}
\end{subequations}
and from Eq.~(\ref{EulerLinearized}) one obtains
\begin{equation}
 \delta^{(1)} \! U = \frac16 \sqrt{\frac{5}{\pi}} \frac{\alpha\Omega^2}{R}
 r^3 \sin^2\theta \cos\theta
 e^{i(2\phi+\omega t)},
 \label{U:noGRforce}
\end{equation}
where $\omega=-4\Omega/3$ is the $l=2$ $r$-mode frequency in the
inertial frame \cite{PapaPringle1978}.

\subsection{Explicit computation of the first-order Eulerian change in the
GR force} \label{EulerVariationGRforce}

To find the linear $r$-mode solution of the perturbed hydrodynamic
equations (\ref{EulerLinearized})--(\ref{PoissonLinearized}) with
the GR reaction force we have to evaluate the first-order Eulerian
change in the GR force, i.e., we have to compute $\delta^{(1)} \!
\vec{F}^{\rm GR}$. To do so we expand both the velocity vector,
$\vec{v}= \hat{\vec{v}}+\delta^{(1)} \! \vec{v}$, and the current
multipole tensors, $S_{ij}= \hat{S}_{ij}+\delta^{(1)} \! S_{ij}$,
where $\hat{S}_{ij}$ describes the multipole tensor of the
unperturbed star and $\delta^{(1)} \! S_{ij}$ is the first-order
Eulerian change in the multipole tensor. The first-order Eulerian
change in the GR force $\delta^{(1)} \! \vec{F}^{\rm GR}$ contains
then terms of the type $\hat{v}_k \delta^{(1)} \! S_{ij}^{[5]}$,
$\hat{S}_{ij}^{[5]} \delta^{(1)} \! v_k$, and $\delta^{(1)} \!
S_{ij}^{[6]}$.

The multipole tensors of the unperturbed star are easily computed
from Eqs.~(\ref{SijSpherical}), with the use of
Eq.~(\ref{UnperturbVeloc}), yielding
\begin{equation}
 \hat{S}_{ij} = 0 \, ; \qquad  i,j=x,y,z\,,
 \label{S:unperturbed}
\end{equation}
i.e., they all vanish. In practice, this result allows us to
rewrite the first-order Eulerian change in the GR force simply as
a sum of terms of the type $\hat{v}_k \delta^{(1)} \!
S_{ij}^{[5]}$ and $\delta^{(1)} \! S_{ij}^{[6]}$.

Now, to compute the first-order Eulerian change in the multipole
tensors $\delta^{(1)} \! S_{ij}$ we use an approach in which we
assume that the linear $r$-mode solution with $\delta^{(1)} \!
\vec{F}^{\rm GR}=0$ acts as a source for the current multipole
tensor. In other words, and taking $\delta^{(1)} \! S_{xx}$ as an
example, we assume that
\begin{eqnarray}
\delta^{(1)} \! S_{xx} &=& -\int dV \,r^2
\sin\theta \cos\phi \nonumber \\
 & & \times {\biggl [}
  \delta^{(1)} \! \rho {\bigl (} \hat{v}_{\theta}\sin\phi +
  \hat{v}_{\phi}\cos\theta \cos\phi
 {\bigr )}  \nonumber \\
 & & + \,\hat{\rho} {\bigl (} \delta^{(1)} \!
 v_{\theta}\sin\phi +\delta^{(1)} \! v_{\phi}\cos\theta
\cos\phi {\bigr )}{\biggr ]},
 \label{deltaS:total}
\end{eqnarray}
where $\hat{\vec{v}}$ is given by Eq.~(\ref{UnperturbVeloc}) and
$\delta^{(1)} \! \vec{v}$ is given by
Eqs.~(\ref{r-mode:noGRforce}) with $\omega$ being now an arbitrary
parameter to be determined. Since $\delta^{(1)} \! \rho$ is of
order $\alpha \Omega^2$, $\delta^{(1)} \! \vec{v}$ is of order
$\alpha \Omega$, and $\hat{\vec{v}}$ is of order $\Omega$, in
Eq.~(\ref{deltaS:total}) there are terms of order $\alpha \Omega$,
namely, $\hat{\rho}\delta^{(1)} v_i$, and terms of order $\alpha
\Omega^3$, namely, $\hat{v}_i \delta^{(1)} \! \rho$. Since our
interest is to find the $r$-mode solution induced by the GR force
to lowest order in $\alpha$ and in $\Omega$, we will neglect in
Eq.~(\ref{deltaS:total}) the contribution coming from the terms
$\hat{v}_i \delta^{(1)} \! \rho$ and we will work only with the
dominant terms $\hat{\rho}\delta^{(1)} \! v_i$. This discussion
applies similarly to the other current multipole tensors, whose
first-order Eulerian change can be straightforwardly written from
Eqs.~(\ref{SijSpherical}).

Under the above conditions, Eqs.~(\ref{r-mode:noGRforce}) are
inserted into Eq.~(\ref{deltaS:total}) yielding
\begin{equation}
 \delta^{(1)} \! S_{xx} = -\alpha \Omega \sqrt{\frac{\pi}{5}}\, \frac{\tilde{J}}{R}
\, e^{i\omega t} + {\cal O} (\alpha \Omega^3),
 \label{deltaS:end}
\end{equation}
where $\omega$ is an arbitrary parameter, ${\cal O} (\alpha
\Omega^3)$ denotes terms of order $\alpha \Omega^3$ or more, and
$\tilde{J}$ is defined as
\begin{equation}
 \tilde{J} \equiv \int_0^R dr \hat{\rho} r^6.
 \label{DefJtilde}
\end{equation}
Similarly, it is straightforward to find that the first-order
Eulerian change in the other multipole tensors satisfy the
relations
\begin{eqnarray}
& &  \delta^{(1)} \! S_{yy} = -\delta^{(1)} \! S_{xx} , \nonumber \\
& &  \delta^{(1)} \! S_{xy} = i\delta^{(1)} \! S_{xx} ,\nonumber \\
& &  \delta^{(1)} \! S_{xz} = \delta^{(1)} \! S_{yz} =
\delta^{(1)} \! S_{zz} = 0.
 \label{deltaSij:all}
\end{eqnarray}
From Eqs.~(\ref{deltaS:end}) and (\ref{deltaSij:all}) it is then
clear that the $n^{\rm th}$ time derivative of the multipole
tensor perturbation is given by
\begin{equation}
 \delta^{(1)} \! S_{ij}^{[n]} = (i \omega)^n \delta^{(1)} \! S_{ij}.
 \label{deltaS:Derivative}
\end{equation}
In spite of the large number of terms that appear in the
definition of the GR force [see Eqs.~(\ref{ForceSpheric}) in the
Appendix~\ref{GRforce}], Eqs.~(\ref{deltaS:end}),
(\ref{deltaSij:all}), and (\ref{deltaS:Derivative}) allow a
considerable simplification of the first-order Eulerian change in
the GR force. Indeed, to lowest order in $\Omega$, the first-order
Eulerian change in the GR force is given simply by
\begin{subequations}
 \label{deltaForce}
\begin{eqnarray}
 \hspace{-0.2cm} \delta^{(1)} \! F_{r}^{\rm GR} &\simeq& -3 i \kappa
 \frac{\tilde{J}}{R} \alpha \Omega^2 \omega^5 r^2 \sin^2\theta
 \cos\theta e^{i(2\phi+\omega t)},
\\
 \hspace{-0.2cm} \delta^{(1)} \! F_{\theta}^{\rm GR} &\simeq& i \kappa
 \frac{\tilde{J}}{R} \alpha \Omega \omega^5 \left(  \omega + 3 \Omega \sin^2\theta
 \right) r^2 \sin\theta e^{i(2\phi+\omega t)},
 \nonumber
\\
\\
 \hspace{-0.2cm} \delta^{(1)} \! F_{\phi}^{\rm GR} &\simeq& -\kappa
 \frac{\tilde{J}}{R} \alpha \Omega \omega^6 r^2 \sin\theta \cos\theta
 e^{i(2\phi+\omega t)},
\end{eqnarray}
\end{subequations}
where the constant $\kappa$ sets the strength of the GR reaction
force and is defined as
\begin{equation}
 \kappa \equiv \frac{16}{45} \sqrt{\frac{\pi}{5}} \,\frac{G}{c^7} .
 \label{Def:kappa}
\end{equation}

\subsection{Linear $\bm r$-mode solution
with the GR reaction contribution}
 \label{Linear r-mode GRforce}

We have now all that is needed to find the linear $r$-mode
solution that solves, to lowest order in $\Omega$, the linearized
fluid equations (\ref{EulerLinearized})--(\ref{PoissonLinearized})
with the GR force.

It is well established that the GR drives the $r$-modes unstable. This means
that the GR reaction induces a small complex imaginary part in the allowed
$r$-mode frequency, i.e., one has
\begin{equation}
\omega = \omega_0 + i \varpi, \label{Def:omegaTotal}
\end{equation}
where $\omega_0\equiv {\rm Re}[\omega]$ is the frequency of the $r$-mode and
$\varpi={\rm Im}[\omega]<0$ is the small imaginary part that is related to the
growth timescale of the instability of the mode.

Assuming that $\varpi/\omega_0\ll 1$, an assumption that will be checked at
the end, the perturbed GR force (\ref{deltaForce}) can finally be written as
\begin{subequations}
 \label{deltaForce:end}
\begin{eqnarray}
& & \hspace{-0.8 cm} \delta^{(1)} \! F_{r}^{\rm GR}  \simeq  - 3 i \kappa
\frac{\tilde{J}}{R} \alpha \Omega^2 \omega_0^5 r^2 \sin^2\theta \cos\theta
e^{-\varpi t} e^{i(2\phi + \omega_0 t)}, \nonumber \\
\\
& & \hspace{-0.8 cm} \delta^{(1)} \! F_{\theta}^{\rm GR} \simeq i
\kappa \frac{\tilde{J}}{R} \alpha \Omega \omega_0^5 \left(
\omega_0 + 3\Omega \sin^2 \theta \right) r^2 \sin\theta  \nonumber
\\
& & \hspace{0.8 cm} \times \, e^{-\varpi t} e^{i(2\phi+\omega_0
t)},
\\
& & \hspace{-0.8 cm} \delta^{(1)} \! F_{\phi}^{\rm GR} \simeq -\kappa
\frac{\tilde{J}}{R} \alpha \Omega \omega_0^6  r^2 \sin\theta \cos\theta
e^{-\varpi t} e^{i(2\phi+\omega_0 t)}.
\end{eqnarray}
\end{subequations}

The perturbed Euler equation (\ref{EulerLinearized}), in the
inertial frame, is given by
\begin{subequations}
 \label{LinearEuler:3eqs}
\begin{eqnarray}
& & \hspace{-0.4 cm}(\partial_t + \Omega \partial_{\phi})
\delta^{(1)} \! v_r - 2 \Omega \sin\theta \delta^{(1)} \! v_{\phi}
+
\partial_r \delta^{(1)} \! U = \delta^{(1)} \! F_r^{\rm GR},
\nonumber
\\
\\
& & \hspace{-0.4 cm} (\partial_t + \Omega \partial_{\phi})
\delta^{(1)} \! v_{\theta} - 2 \Omega \cos\theta \delta^{(1)} \!
v_{\phi} + \frac{1}{r}
\partial_{\theta} \delta^{(1)} \! U
= \delta^{(1)} \! F_{\theta}^{\rm GR}, \nonumber
\\
\\
& & \hspace{-0.4 cm} (\partial_t + \Omega \partial_{\phi})
\delta^{(1)} \! v_{\phi} + 2 \Omega \sin\theta \delta^{(1)} \!
v_{r} + 2 \Omega \cos\theta \delta^{(1)} \! v_{\theta} \nonumber
\\
& & \hspace{1 cm} + \, \frac{1}{r\sin\theta} \partial_{\phi}
\delta^{(1)} \! U = \delta^{(1)} \! F_{\phi}^{\rm GR},
\end{eqnarray}
\end{subequations}
with $\delta^{(1)} \! \vec{F}^{\rm GR}$ defined by
Eqs.~(\ref{deltaForce:end}). In order to solve
Eqs.~(\ref{LinearEuler:3eqs}), we try a solution that has the same
structure as the GR force-free $r$-mode solution given by
Eqs.~(\ref{r-mode:noGRforce}) and (\ref{U:noGRforce}). More
specifically, we try an {\it Ansatz} of the type\footnote{At a
first glance we might expect that the GR $r$-mode solution should
have a non-vanishing radial velocity component since the
first-order Eulerian change in the GR force (\ref{deltaForce:end})
has a non-vanishing radial component. However, this is not
necessarily the result that we might expect, as the experience
acquired with the GR force-free $r$-mode indicates. Indeed, in
this case there is no GR force, but the Coriolis force, which is
the restoring force, is present. Now, although the Coriolis force
has a radial component, the fact is that, to leading order, the
force-free $r$-mode velocity perturbations
(\ref{r-mode:noGRforce}) do not have a radial component.}
\begin{subequations}
 \label{r-mode:GR:try}
\begin{eqnarray}
\delta^{(1)} \! v_r &=& 0,
\\
\delta^{(1)} \! v_{\theta} &=& -\frac{i}{4} \sqrt{\frac{5}{\pi}}
\frac{\alpha \Omega}{R} r^2 \sin\theta \,e^{-\varpi t}
e^{i(2\phi+\omega_0 t)} \nonumber
\\
& & + \, \delta^{(1)} \! \tilde{v}_{\theta}(t,r,\theta,\phi),
\\
\delta^{(1)} \! v_{\phi} &=& \frac14 \sqrt{\frac{5}{\pi}}
\frac{\alpha \Omega}{R} r^2 \sin\theta \cos\theta \,e^{-\varpi t}
e^{i(2\phi + \omega_0 t)} \nonumber
\\
& & + \,  \delta^{(1)} \! \tilde{v}_{\phi}(t,r,\theta,\phi),
\end{eqnarray}
\end{subequations}
and
\begin{eqnarray}
\delta^{(1)} \! U &=& \frac16 \sqrt{\frac{5}{\pi}} \frac{\alpha
\Omega^2}{R} r^3 \sin^2\theta \cos\theta \,e^{-\varpi t}
e^{i(2\phi+\omega_0 t)} \nonumber
\\
& & + \, \delta^{(1)} \! \tilde{U} (t,r,\theta,\phi),
 \label{r-mode:GR:try:U}
\end{eqnarray}
where $\delta^{(1)} \! \tilde{\vec{v}}$ and $\delta^{(1)} \!
\tilde{U}$ are functions of $t$, $r$, $\theta$ and $\phi$ to be
determined, and $\varpi$ and $\omega_0$ are unknown constants to
be fixed.

As already mentioned above [see
Eq.~(\ref{GRtimescale:LindOwenMors})], it is known from previous
investigations \cite{LindOwenMors1998,AndersKokkSchutz1999} on the
time evolution of the physical energy of the $r$-mode that
$\varpi=1/\tau_{GR}$ is proportional to $\Omega^6$. This result
follows straightforwardly from an inspection of
Eqs.~(\ref{LinearEuler:3eqs}) with $\delta^{(1)} \! \vec{F}^{\rm
GR}$ and $\delta^{(1)} \! \vec{v}$ defined by
Eqs.~(\ref{deltaForce:end}) and (\ref{r-mode:GR:try}),
respectively. Indeed, the time derivative of $\delta^{(1)} \!
\vec{v}$ on the left-hand side of Eqs.~(\ref{LinearEuler:3eqs})
yields terms proportional to $\Omega\varpi$, while the right-hand
side of these equations contains, to leading order in $\Omega$,
terms proportional to $\Omega^7$. Similarly, the $r$-mode
frequency $\omega_0$ could also have a correction (relatively to
the GR force-free value $\omega_0=-4\Omega/3$) proportional to
$\Omega^6$. Therefore, we write $\varpi$ and $\omega_0$ in the
form
\begin{equation}
\varpi =- \frac{8}{9}\sqrt{\frac{\pi}{5}}\, \gamma R \Omega^6 a
 \label{FixingSigma}
\end{equation}
and
\begin{equation}
 \omega_0 = - \frac{4\Omega}{3}+\frac89\sqrt{\frac{\pi}{5}}\gamma R \Omega^6
 b,
 \label{Def:omega0}
\end{equation}
where $a$ and $b$ are (by now arbitrary) real dimensionless
constants and we have defined
\begin{equation}
 \gamma \equiv \frac{2^{10}}{3^4} \,
 \frac{\tilde{J}}{R} \, \kappa.
 \label{Def:gamma}
\end{equation}
Note that for $a=1$ the growth timescale of the GR-induced
$r$-mode instability, $\tau_{GR}=1/\varpi$, coincides with the one
obtained in Refs.~\cite{LindOwenMors1998,AndersKokkSchutz1999}.
For $b=0$, the frequency $\omega_0$ coincides with that of the GR
force-free linear $r$-mode solution.

For such an {\it Ansatz} the perturbed Euler equation reduces to
the following system of differential equations for $\delta^{(1)}
\! \tilde{v}_\theta$, $\delta^{(1)} \! \tilde{v}_\phi$ and
$\delta^{(1)} \! \tilde{U}$,
\begin{subequations}
 \label{Eqs:Vtilde}
\begin{eqnarray}
& & \hspace{-0.5 cm} - 2 \Omega \sin\theta \delta^{(1)} \!
\tilde{v}_{\phi} +
\partial_r \delta^{(1)} \! \tilde{U} \nonumber \\
& &  = i \gamma \alpha \Omega^7 r^2 \sin^2\theta \cos\theta
e^{-\varpi t} e^{i(2\phi+\omega_0 t)}, \label{Eqs:Vtilde-r}
\\
& & \hspace{-0.5 cm}  ( \partial_t + \Omega \partial_{\phi} )
\delta^{(1)} \! \tilde{v}_{\theta} - 2 \Omega \cos\theta
\delta^{(1)} \! \tilde{v}_{\phi} + \frac{1}{r}
\partial_{\theta} \delta^{(1)} \! \tilde{U} \nonumber
\\ & & = i \gamma \left[ \frac29
(a+bi+2) - \sin^2\theta \right] \alpha \Omega^7 r^2
\sin\theta \nonumber \\
&& \hspace{0.3 cm} \times \, e^{-\varpi t} e^{i(2\phi+\omega_0
t)},
\\
& & \hspace{-0.5cm} ( \partial_t + \Omega \partial_{\phi})
\delta^{(1)} \! \tilde{v}_{\phi} + 2 \Omega \cos\theta
\delta^{(1)} \! \tilde{v}_{\theta} + \frac{1}{r\sin\theta}
\partial_{\phi} \delta^{(1)} \! \tilde{U} \nonumber
\\
& &  = - \frac{2\gamma}{9} \left( a+bi+2 \right) \alpha \Omega^7
r^2 \sin\theta \cos\theta e^{-\varpi t}
e^{i(2\phi+\omega_0 t)}. \nonumber \\
\end{eqnarray}
\end{subequations}

The right-hand side of the above system induces a solution of the
form
\begin{subequations}
 \label{tildev}
\begin{eqnarray}
\delta^{(1)} \! \tilde{v}_{\theta} &=& i \gamma \alpha \Omega^6
r^2 f(\theta) \,e^{-\varpi t} e^{i(2\phi+\omega_0 t)},
\\
\delta^{(1)} \! \tilde{v}_{\phi} &=& -\gamma \alpha \Omega^6 r^2
g(\theta) \,e^{-\varpi t} e^{i(2\phi+\omega_0 t)},
\end{eqnarray}
\end{subequations}
and
\begin{eqnarray}
\delta^{(1)} \! \tilde{U} &=& \frac{i}{3} \gamma \alpha \Omega^7
r^3 h(\theta) \,e^{-\varpi t} e^{i(2\phi+\omega_0 t)},
 \label{tildeU}
\end{eqnarray}
where $f(\theta)$, $g(\theta)$ and $h(\theta)$ are functions of
$\theta$ determined by the system of equations
\begin{subequations}
 \label{set3equations}
\begin{eqnarray}
& & 2\sin\theta g(\theta) +i h(\theta) = i \sin^2\theta
\cos\theta,
\\
& & 2 f(\theta) - 6\cos\theta g(\theta) -i
\partial_{\theta} h(\theta) \nonumber \\
& & \hspace{0.8 cm} = -3i \sin\theta \left[ \frac29(a+bi+2)
-\sin^2\theta \right],
\\
& & 3i \sin\theta \cos\theta f(\theta) -i \sin\theta g(\theta)
-h(\theta) \nonumber \\
& & \hspace{0.8 cm} = -\frac13 \left( a+bi+2 \right) \sin^2\theta
\cos\theta.
\end{eqnarray}
\end{subequations}
From the system of equations (\ref{set3equations}) it is
straightforward to obtain a linear first-order differential
equation for $h(\theta)$, namely,
\begin{equation}
\frac{dh (\theta)}{d\theta}+ \frac{1-3\cos^2\theta}{\sin\theta
\cos\theta} h(\theta)=q(\theta), \label{eq:h}
\end{equation}
where the source term $q(\theta)$ is given by
\begin{equation}
q(\theta) = \frac89 \left( a+bi-1 \right) \sin\theta.
\label{q-theta}
\end{equation}
This equation has the solution
\begin{eqnarray}
h(\theta) &=& A\sin^2\theta \cos\theta \nonumber \\
& & + \, \frac89 (a+bi-1) \sin^2\theta \cos\theta \ln(\tan\theta),
\label{h-theta}
\end{eqnarray}
where $A$ is a (complex) constant determined by initial data. In
Eq.~(\ref{h-theta}), the first term of the right-hand side
corresponds to the general solution of the homogeneous equation
[Eq.~(\ref{eq:h}) with $q(\theta)=0$], and the second term on the
right-hand side corresponds to the particular solution of
Eq.~(\ref{eq:h}) with the source term $q(\theta)$. From
Eqs.~(\ref{set3equations}) and (\ref{h-theta}) is then
straightforward to obtain:
\begin{eqnarray}
f(\theta)&=& -\frac{i}{2}(A-1) \sin\theta + \frac{i}{9}
(a+bi-1)\sin\theta \nonumber \\
&& -\frac{4i}{9} (a+bi-1) \sin\theta \ln(\tan\theta) , \label{f}
\\
g(\theta)&=&-\frac{i}{2} (A-1) \sin\theta\cos\theta \nonumber
\\
&& -\frac{4i}{9} (a+bi-1) \sin\theta\cos\theta
\ln(\tan\theta).\label{g}
\end{eqnarray}

The above solution to the perturbed Euler equation must also
satisfy the continuity equation (\ref{ContinuityLinearized}),
which at order $\alpha\Omega^6$ is simply given by
\begin{equation}
\vec{\nabla} \cdot \delta^{(1)}\tilde{\vec{v}}=0. \label{cont}
\end{equation}
Inserting $\delta^{(1)}\tilde{\vec{v}}$ from Eq.~(\ref{tildev}),
with $f(\theta)$ and $g(\theta)$ given by Eqs.~(\ref{f}) and
(\ref{g}), into the continuity equation (\ref{cont}) we obtain:
\begin{eqnarray}
\partial_\theta \left[ \sin\theta f(\theta) \right] -
2 g(\theta)=0
\end{eqnarray}
or
\begin{eqnarray}
\hspace{-0.2 cm} -\frac{4i}{9} (a+bi-1)
\frac{\sin\theta}{\cos\theta} + \frac{2i}{9} (a+bi-1) \sin\theta
\cos\theta=0,
\end{eqnarray}
implying that $a=1$ and $b=0$.

Therefore, we arrive at the conclusion that the $r$-mode velocity
perturbations are sinusoidal with the same frequency as the
well-known GR force-free linear $r$-mode solution, and that the GR
force drives the $r$-modes unstable with a growth timescale that
agrees with the expression found in
Refs.~\cite{LindOwenMors1998,AndersKokkSchutz1999}.

Inserting solutions (\ref{tildev}) and (\ref{tildeU}) [with
$f(\theta)$, $g(\theta)$ and $h(\theta)$ given by
Eqs.~(\ref{h-theta})--(\ref{g}), $a=1$ and $b=0$] into
Eqs.~(\ref{r-mode:GR:try}) and (\ref{r-mode:GR:try:U}) we finally
get that the $r$-mode solution to the linearized fluid equations
with the GR force is given by
\begin{subequations}
 \label{r-mode:GR}
\begin{eqnarray}
\delta^{(1)} \! v_r &=& 0,
\\
\delta^{(1)} \! v_{\theta} &=& - \frac{i}{2} \alpha \Omega \left[
\frac12 \sqrt{\frac{5}{\pi}} \frac{1}{R} + i \gamma (A-1) \Omega^5
\right] r^2 \sin\theta \nonumber
\\
& & \times \, e^{-\varpi t} e^{i(2\phi+\omega_0 t)},
\\
\delta^{(1)} \! v_{\phi} &=& \frac12 \alpha \Omega \left[ \frac12
\sqrt{\frac{5}{\pi}} \frac{1}{R} + i \gamma (A-1) \Omega^5 \right]
r^2 \sin\theta \cos\theta
 \nonumber
\\
& & \times \, e^{-\varpi t} e^{i(2\phi+\omega_0 t)},
\end{eqnarray}
\end{subequations}
and
\begin{eqnarray}
\delta^{(1)} \! U &=& \frac{1}{3} \alpha \Omega^2 \left[ \frac12
\sqrt{\frac{5}{\pi}} \frac{1}{R} +i \gamma A \Omega^5 \right] r^3
\sin^2\theta \cos\theta
\nonumber \\
& & \times \, e^{-\varpi t}
  e^{i(2\phi+\omega_0 t)},
 \label{r-mode:GR:U}
\end{eqnarray}
with $\varpi$ and $\omega_0$ given by
\begin{equation}
\varpi =- \frac{8}{9}\sqrt{\frac{\pi}{5}}\, \gamma R \Omega^6
 \label{FixingSigma-final}
\end{equation}
and
\begin{equation}
 \omega_0 = - \frac{4\Omega}{3}.
 \label{Def:omega0-final}
\end{equation}

The velocity perturbations given by Eqs.~(\ref{r-mode:GR}) have a
piece similar to the GR force-free solution, the difference being
the factor $e^{-\varpi t}$ responsible for the exponential growth
of the $r$-mode amplitude due to the presence of a GR reaction
force, and another piece proportional to $\alpha\gamma (A-1)
\Omega^6$, where $A$ is a constant fixed by the choice of initial
data. If this constant is chosen to be of order unity, $A\sim
\mathcal{O}(1)$, then the second term in the square brackets of
Eqs.~(\ref{r-mode:GR}) and (\ref{r-mode:GR:U}) is much smaller
than the first term. Indeed, let us assume that the mass density
$\rho$ and the pressure $P$ \! of the perfect-fluid star are
related by the usual polytropic equation of state $P=k\rho^2$,
with $k$ such that $M=1.4 M_{\bigodot}$ and $R=12.53$ km. For such
a choice of the equation of state, $\tilde{J}=1.12\times10^{45}
\mbox{ kg m}^4$ and the Keplerian angular velocity at which the
star starts shedding mass through the equator is
$\Omega_K=\frac23\sqrt{\pi G \bar{\rho}}=5612 \mbox{ sec}^{-1}$,
where $\bar{\rho}=3.38\times10^{17}\mbox{ kg/m}^3$ is the average
mass density of the star and $G$ is the gravitational constant.
Then, in Eqs.~(\ref{r-mode:GR}) [or similarly in
Eq.~(\ref{r-mode:GR:U})] the expression inside the square brackets
can be written as $\sqrt{5/(4\pi R^2)}[1+1.08\times10^{-5} i (A-1)
(\Omega/\Omega_K)^5]$. For $A\sim \mathcal{O}(1)$ and
$\Omega\leqslant \Omega_K$, the second term in the square bracket
can be neglected.

Note that in the GR force-free limit, the constant $\kappa$ that
sets the strengh of the GR reaction force goes to zero. Therefore,
in this limit $\gamma$ defined in Eq.~(\ref{Def:gamma}) goes to
zero and the  $\delta^{(1)} \! \tilde{\vec{v}}$ perturbations
vanish as they should, since they are proportional to $\gamma$.

\section{Discussion of the results} \label{Conclusion}

Analytical $r$-mode solutions were investigated, within the
linearized theory, in the case of a slowly rotating, Newtonian,
barotropic, non-magnetized, perfect-fluid star in which a
gravitational radiation (GR) reaction force is present. For the GR
reaction term we have used the 3.5 post-Newtonian order expansion
of the GR force, in order to include the contribution of the
current quadrupole moment. The evaluation of the GR force
perturbation requires the knowledge of the multipole moment
perturbation and its time derivatives. To compute this Eulerian
change in the multipole moment we have used an approach in which
we assumed that the GR force-free linear $r$-mode solution acts as
a source for the current multipole moment. We have carried our
analysis of the perturbed fluid equations only to lowest order in
$\Omega$. This order is enough to evaluate the GR reaction
contribution. A higher order analysis would include perturbations
in the density which would be important if we were interested in
the accurate evaluation of the viscosity effects.

We have found the analytical expression of the $r$-mode velocity
perturbations that solves the Newtonian hydrodynamic equations
with the GR reaction force. From this solution, given by
Eqs.~(\ref{r-mode:GR}) and (\ref{r-mode:GR:U}), four important
features of the $r$-mode instability driven by the GR reaction can
be read:

\noindent (i) The velocity perturbations $\delta^{(1)} \! \vec{v}$
are proportional to $e^{i(2\phi+\omega_0 t)}$, with
$\omega_0=-4\Omega/3$. Thus, they have the same sinusoidal
behavior and the same frequency $\omega_0$ as the solution of the
GR force-free linear velocity perturbations given by
Eqs.~(\ref{r-mode:noGRforce}).

\noindent (ii) The amplitude of the velocity perturbations is proportional to
$\exp\{-\varpi t\}$. Since $\varpi<0$, the GR force induces then an
exponential growth in the $r$-mode amplitude. The e-folding growth timescale
$\tau_{\rm GR}=1/\varpi$ agrees with the GR timescale
(\ref{GRtimescale:LindOwenMors}) first found in Ref.~\cite{LindOwenMors1998}.
Thus, these velocity perturbations are consistent with the expression found in
Ref.~\cite{LindOwenMors1998} for the time evolution of the energy of the
$r$-mode perturbation, $dE/dt$.

\noindent (iii) The velocity perturbations $\delta^{(1)} \!
\vec{v}$ contain a piece proportional to $\gamma (A-1) \alpha
\Omega^6$, where $A$ is an arbitrary constant fixed by the choice
of initial data. If we choose this constant $A$ to be of order
unity, then this part of the solution could be neglected.

\noindent (iv) The parameter $\kappa$, defined in
Eq.~(\ref{Def:kappa}), sets the strength of the GR reaction force.
Thus, the GR force-free limit is obtained when we set $\kappa$
equal to zero. In this limit, $\varpi$ and $\gamma$ defined,
respectively, by Eqs.~(\ref{FixingSigma}) and (\ref{Def:gamma}) go
to zero. Then, from Eqs.~(\ref{r-mode:GR}) and (\ref{r-mode:GR:U})
we recover the GR force-free linear $r$-mode solution given by
Eqs.~(\ref{r-mode:noGRforce}) and (\ref{U:noGRforce}).

Our results are strictly valid only for the slow-rotation regime
since our solution holds only to leading order. However, there are
good indications that the evolution of $r$-modes in
rapidly-spinning stars does not differ significantly from the
evolution in slowly-rotating stars \cite{Yoshida2000}. Hence, it
is reasonable to expect that at least the qualitative features
that we found are also valid in the rapid-rotation regime.

A natural extension of the present work is to try to find an
analytical $r$-mode solution of the nonlinear hydrodynamic
equations with the GR reaction force. The nonlinear GR force-free
case was analyzed in Ref.~\cite{Sa2004}, where an analytical
$r$-mode solution in the nonlinear theory up to second order in
the mode amplitude was found. This solution represents
differential rotation that produces large scale drifts of fluid
elements along stellar latitudes. It contains two separate pieces,
one induced by first-order quantities and another determined by
the choice of initial data. Since these two pieces cannot cancel
each other, differential rotation is an unavoidable nonlinear
kinematic feature of $r$-modes. The analysis of Ref.~\cite{Sa2004}
can be extended to the case in which the GR reaction force is
present. The main purpose of this extension is to find if the GR
reaction force provides an extra source of differential rotation.
This work is in progress \cite{DiasSa}.

\section*{Acknowledgements}
It is a pleasure to acknowledge Shijun Yoshida, Kostas Kokkotas
and Luciano Rezzolla for useful suggestions that improved the
manuscript. We also acknowledge the anonymous referee for her/his
valuable comments. This work was supported in part by Funda\c
c\~ao para a Ci\^encia e Tecnologia (FCT). OJCD acknowledges
financial support from FCT through grant SFRH/BPD/2004.



\appendix
\section{Explicit computation of the GR reaction force}
\label{GRforce}

In the cartesian basis the explicit components of the
gravitational vector potential (\ref{DefBeta}) are
\begin{subequations}
\label{betaCartesian:Sij}
\begin{eqnarray}
\beta_x &=& \frac{16G}{45c^7} \left[ y z \left( S_{zz}^{[5]} -
S_{yy}^{[5]} \right) \right. \nonumber
\\
& &  \left. -\, x z S_{xy}^{[5]} + x y S_{xz}^{[5]} + (y^2-z^2)
S_{yz}^{[5]} \right],
\\
\beta_y &=& \frac{16G}{45c^7} \left[ x z \left( S_{xx}^{[5]} -
S_{zz}^{[5]} \right) \right. \nonumber
\\
& & \left. +\, y z S_{xy}^{[5]} - (x^2-z^2) S_{xz}^{[5]} - x y
S_{yz}^{[5]} \right],
\\
\beta_z &=& \frac{16G}{45c^7} \left[ x y \left( S_{yy}^{[5]} -
S_{xx}^{[5]} \right) \right. \nonumber
\\
& & \left. +\, (x^2-y^2) S_{xy}^{[5]} - y z S_{xz}^{[5]} + x z
S_{yz}^{[5]} \right].
\end{eqnarray}
\end{subequations}
The use of the usual relations between cartesian  $(x,y,z)$ and
spherical $(r,\theta,\phi)$ coordinates yields the components of
the gravitational vector potential $\vec{\beta}$ in spherical
coordinates,
\begin{subequations}
 \label{betaSpherical:J22}
\begin{eqnarray}
& &  \hspace{-0.4cm} \beta_r = 0,
\\
& &  \hspace{-0.4cm} \beta_{\theta} = -\frac{16G}{45c^7} r^2
{\biggl \{} \cos\theta \left( \cos\phi S_{yz}^{[5]}-\sin\phi
S_{xz}^{[5]} \right) \nonumber \\
& &  \hspace{-0.3 cm} + \, \sin\theta \left[ \cos(2\phi)
S_{xy}^{[5]} + \frac{1}{2} \sin(2\phi) \left(
S_{yy}^{[5]}-S_{xx}^{[5]} \right) \right] {\biggr \}},
\\
& &
\hspace{-0.4cm} \beta_{\phi} = \frac{8G}{45c^7} r^2
{\biggl \{} 2 \cos(2\theta) \cos\phi S_{xz}^{[5]} \nonumber \\
& &  \hspace{-0.3 cm} +\,  \frac{1}{2} \sin(2\theta) \left[
S_{xx}^{[5]} + S_{yy}^{[5]} - 2S_{zz}^{[5]} + \cos(2\phi) \left(
S_{xx}^{[5]} - S_{yy}^{[5]} \right) \right]
  \nonumber \\
& & \hspace{-0.3 cm} + \, 2\cos(2\theta) \sin\phi S_{yz}^{[5]} +
\sin(2\theta) \sin(2\phi) S_{xy}^{[5]} {\biggr \}},
\end{eqnarray}
\end{subequations}
where the current multipole tensor in spherical coordinates is
given explicitly by
\begin{subequations}
\label{SijSpherical}
\begin{eqnarray}
& & \hspace{-0.4cm} S_{xx} = -\int dV \rho r^2 \sin\theta \cos\phi
\left( v_{\theta}\sin\phi + v_{\phi}\cos\theta \cos\phi \right),
\nonumber
\\
\end{eqnarray}
\begin{eqnarray}
& & \hspace{-0.4cm} S_{yy} = \int dV \rho r^2 \sin\theta \sin\phi
\left( v_{\theta}\cos\phi -v_{\phi}\cos\theta \sin\phi
 \right), \nonumber
\\
\\
& & \hspace{-0.4cm} S_{zz} = \int dV \rho r^2 v_{\phi} \sin\theta
\cos\theta,
\\
& & \hspace{-0.4cm} S_{xy}= \frac14 \int dV \rho r^2
\left[ v_{\theta} \sin\theta \cos(2\phi) - v_{\phi}\sin(2\theta)
\sin(2\phi) \right], \nonumber
\\
\\
& & \hspace{-0.4cm} S_{xz} = - \frac12 \int dV \rho r^2 \left[
v_{\theta}\cos\theta\sin\phi + v_{\phi}\cos(2\theta) \cos\phi
\right], \nonumber
\\
\\
& & \hspace{-0.4cm} S_{yz} = \frac12 \int dV \rho r^2 \left[
v_{\theta}\cos\theta\cos\phi - v_{\phi}\cos(2\theta) \sin\phi
\right]\! .
\end{eqnarray}
\end{subequations}

Inserting Eqs.~(\ref{betaSpherical:J22}) into Eq.~(\ref{DefGravForce}) yields
the spherical components of the GR reaction force,
\begin{widetext}
\begin{subequations}
 \label{ForceSpheric}
\begin{eqnarray}
\hspace{-0.6cm} F_{r}^{\rm GR} &=& -\frac{16G}{15c^7} r v_{\theta}
\left\{ \cos\theta \left( \cos\phi S_{yz}^{[5]} - \sin\phi
S_{xz}^{[5]} \right) + \sin\theta \left[ \frac{1}{2} \sin(2\phi)
\left( -S_{xx}^{[5]}+S_{yy}^{[5]} \right) + \cos(2\phi)
S_{xy}^{[5]} \right] \right\} \nonumber
\\
& & + \frac{8G}{15c^7} r v_{\phi} {\biggl \{} \sin(2\theta)
\sin(2\phi) S_{xy}^{[5]} + \frac{1}{2} \sin(2\theta) \left[ 3
\left( S_{xx}^{[5]}+S_{yy}^{[5]} \right) + \cos(2\phi) \left(
S_{xx}^{[5]}-S_{yy}^{[5]} \right) \right] \nonumber
\\
& & \hspace{1.8cm} +2 \cos(2\theta) \left( \cos\phi
S_{xz}^{[5]}+\sin\phi S_{yz}^{[5]} \right) {\biggr \}},
\end{eqnarray}
\begin{eqnarray}
\hspace{-0.6cm} F_{\theta}^{\rm GR} &=& \frac{16G}{45c^7} r^2
\left\{ \cos\theta \left( \cos\phi S_{yz}^{[6]} - \sin\phi
S_{xz}^{[6]} \right) + \sin\theta \left[ \cos(2\phi) S_{xy}^{[6]}
+ \frac{1}{2} \sin(2\phi) \left( S_{yy}^{[6]} - S_{xx}^{[6]}
\right) \right] \right\} \nonumber
\\
& & + \frac{4G}{45c^7} r v_{\phi} {\biggl \{} 3 \left(
S_{xx}^{[5]} + S_{yy}^{[5]} \right) \left( 1 + 3 \cos(2\theta)
\right) - 6 \sin^2\theta \left[ \cos(2\phi) \left( S_{xx}^{[5]} -
S_{yy}^{[5]} \right) + 2 \sin(2\phi) S_{xy}^{[5]} \right]
\nonumber
\\
& & \hspace{1.8cm} -12 \sin(2\theta) \left( \cos\phi
S_{xz}^{[5]}+\sin\phi S_{yz}^{[5]} \right) {\biggr \}} \nonumber
\\
& & + \frac{16G}{15c^7} r v_{r} \left\{ \cos\theta \left( \cos\phi
S_{yz}^{[5]} - \sin\phi S_{xz}^{[5]} \right) + \sin\theta \left[
\frac{1}{2} \sin(2\phi) \left( -S_{xx}^{[5]} + S_{yy}^{[5]}
\right) + \cos(2\phi) S_{xy}^{[5]} \right] \right\},
\end{eqnarray}
\begin{eqnarray}
\hspace{-0.6cm} F_{\phi}^{\rm GR} &=& - \frac{8G}{45c^7} r^2
{\biggl \{} 2 \cos(2\theta) \cos\phi S_{xz}^{[6]} + \frac{1}{2}
\sin(2\theta) \left[ 3 \left( S_{xx}^{[6]} + S_{yy}^{[6]} \right)
+ \cos(2\phi) \left( S_{xx}^{[6]} - S_{yy}^{[6]} \right) \right]
\nonumber
\\
& & \hspace{1.6cm} + \, 2 \cos(2\theta) \sin\phi S_{yz}^{[6]} +
\sin(2\theta) \sin(2\phi) S_{xy}^{[6]} {\biggr \}} \nonumber
\\
& & - \frac{8G}{15c^7} r v_{r} {\biggl \{} \sin(2\theta)
\sin(2\phi) S_{xy}^{[5]} + \frac{1}{2} \sin(2\theta)  \left[ 3
\left( S_{xx}^{[5]} + S_{yy}^{[5]} \right) + \cos(2\phi) \left(
S_{xx}^{[5]}-S_{yy}^{[5]} \right) \right] \nonumber
\\
& & \hspace{1.8cm} +2 \cos(2\theta) \left( \cos\phi
S_{xz}^{[5]}+\sin\phi S_{yz}^{[5]} \right) {\biggr \}} \nonumber
\\
& & - \frac{4G}{45c^7} r v_{\theta} {\biggl \{} 3 \left(
S_{xx}^{[5]} + S_{yy}^{[5]} \right) \left( 1 + 3 \cos(2\theta)
\right) - 6 \sin^2\theta \left[ \cos(2\phi) \left( S_{xx}^{[5]} -
S_{yy}^{[5]} \right) + 2 \sin(2\phi) S_{xy}^{[5]} \right]
\nonumber
\\
& & \hspace{1.8cm} -12 \sin(2\theta) \left( \cos\phi S_{xz}^{[5]}
+ \sin\phi S_{yz}^{[5]} \right) {\biggr \}},
\end{eqnarray}
\end{subequations}
\end{widetext}
where we have made use of the relation $S_{zz}= -(S_{xx} +S_{yy})$
[see Eqs.~(\ref{SijSpherical})] to remove $S_{zz}$ from the final
expression.

As a check to our expressions for the GR reaction terms, we apply
them to the spherical shell system discussed in detail in
Ref.~\cite{LevinUsh2001}. The authors have computed the
instability timescale driven by the GR reaction and concluded that
it is given by Eq.~(\ref{GRtimescale:LindOwenMors}) with
$\tilde{J}\equiv \rho R^6$, and $\rho$ and $R$ being the surface
density and the radius of the spherical shell. Now, if we
introduce the $r$-mode velocity given by Eqs.~(1) and (7) of
Ref.~\cite{LevinUsh2001} into our Eqs.~(\ref{betaSpherical:J22})
and (\ref{SijSpherical}), we obtain Eqs.~(B14) and (55) of
Ref.~\cite{LevinUsh2001} that define the timescale of the
instability on the shell.

\section{Connection to previous works} \label{Comparing}

In previous works, in which the effects of the GR force on the
$r$-mode evolution were studied numerically, the final form of the
GR force was always written as a function of the multipole moments
$J_{lm}$, instead of the quadrupole tensors $S_{ij}$
\cite{LindTohVal2001,LindTohVal2002,GressmanEtAll2003}. To compare
our expressions with those of
Refs.~\cite{LindTohVal2001,LindTohVal2002,GressmanEtAll2003} we
establish the connection between the two approaches in this
Appendix.

The current multipole moments are defined as \cite{Thorne1980}
\begin{equation}
 J_{lm} = \int dV \rho r^l \vec{v} \cdot \vec{Y}_{lm}^{B\ast},
 \label{DefJlm}
\end{equation}
where $l$ and $m$ are the angular momentum and the azimuthal
numbers, respectively, $\vec{Y}_{lm}^{B\ast}$ is a spherical
harmonic vector of the magnetic type defined by \cite{Thorne1980}
\begin{equation}
 \vec{Y}_{l m}^{B\ast} = (-1)^m \vec{Y}_{l\, -m}^{B}
 \label{DefYlmB*}
\end{equation}
and
\begin{equation}
 \vec{Y}_{lm}^{B} = [l(l+1)]^{-1/2} \vec{r} \times ( \vec{\nabla} Y_{lm}).
 \label{DefYlmB}
\end{equation}
The usual spherical harmonic functions $Y_{l m}$ are given by
\begin{equation}
 Y_{lm} = (-1)^m \left[ \frac{2l+1}{4\pi}\frac{(l-m)!}{(l+m)!} \right]^{1/2}
 e^{i m \phi} P_l^m(\cos\theta),
 \label{DefYlm}
\end{equation}
where $\theta$ and $\phi$ are the angular spherical coordinates
and $P_l^m(\cos \theta)$ are the associated Legendre functions
defined by
\begin{equation}
 P_l^m(\cos\theta) = \frac{1}{2^l l!} \sin^m\theta
 \frac{d^{\,l+m}}{d \cos \theta^{l+m}} (\cos^2 \theta-1)^l.
 \label{DefPlm}
\end{equation}
Like in our case, the authors of
Refs.~\cite{LindTohVal2001,LindTohVal2002,GressmanEtAll2003} are
interested only in the most unstable mode, the $l=2$ $r$-mode.
This mode excites the $J_{2m}$ current multipole moments which in
a spherical coordinate system are written as
\begin{subequations}
 \label{Jspherical}
\begin{eqnarray}
&& \hspace{-0.9cm} J_{22} = \frac{1}{4} \sqrt{\frac{5}{\pi}} \int
dV \rho
 e^{-2i\phi} r^2 \sin\theta \left( v_{\phi} \cos\theta +
 i v_{\theta}\right),
\\
&& \hspace{-0.9cm} J_{21} = -\frac{1}{4} \sqrt{\frac{5}{\pi}} \int
dV \rho e^{-i\phi} r^2 \left[ v_{\phi} \cos(2\theta) + i
v_{\theta}\cos\theta \right], \nonumber
\\
\\
& & \hspace{-0.9cm} J_{20} = -\frac{1}{4} \sqrt{\frac{15}{2\pi}}
\int dV \rho
 r^2 v_{\phi} \sin(2\theta).
\end{eqnarray}
\end{subequations}
From Eqs.~(\ref{SijSpherical}) and (\ref{Jspherical}) it is
straightforward to verify the following correspondence between the
current quadrupole tensor $S_{ij}$ and the current multipole
moments $J_{2m}$ \cite{LindTohVal2002}:
\begin{subequations}
 \label{relation:Sij-J2m}
\begin{eqnarray}
& &  S_{yy} - S_{xx} + 2 i S_{xy} = 4\sqrt{\frac{\pi}{5}} J_{22},
 \label{relation:Sij-J2ma}
\\
& &  S_{xz} - i S_{yz} = 2\sqrt{\frac{\pi}{5}} J_{21},
 \label{relation:Sij-J2mb}
\\
& &  S_{xx} + S_{yy} = - S_{zz} = 2\sqrt{\frac{2\pi}{15}} J_{20}.
 \label{relation:Sij-J2mc}
\end{eqnarray}
\end{subequations}
The most unstable $r$-mode is in fact the $l=m=2$ mode which
excites $J_{22}$ but not $J_{21}$ and $J_{20}$. Hence, in
Refs.~\cite{LindTohVal2001,LindTohVal2002,GressmanEtAll2003}, the
GR reaction force is computed assuming that, in
Eq.~(\ref{relation:Sij-J2m}), $J_{21}\simeq 0$ and $J_{20}\simeq
0$, i.e.,
\begin{subequations}
\label{relation:Sij-J2mv2}
\begin{eqnarray}
& &  S_{yy} - S_{xx} + 2 i S_{xy} = 4\sqrt{\frac{\pi}{5}} J_{22},
\label{relation:Sij-J2mv2a}
\\
& &  S_{xz} - i S_{yz} \simeq 0, \label{relation:Sij-J2mv2b}
\\
& &  S_{xx} + S_{yy} = - S_{zz} \simeq 0.
\label{relation:Sij-J2mv2c}
\end{eqnarray}
\end{subequations}
It is also assumed there that
\begin{equation}
J_{22}^{[n]} \simeq (i \omega)^n  J_{22},
 \label{J22:Derivative}
\end{equation}
and it is found that the GR force in the inertial frame is given
by
\begin{subequations}
 \label{ForceSphericJ22}
\begin{eqnarray}
\hspace{-0.6cm} F_{r}^{\rm GR} &\simeq& 3 \kappa (iv_{\theta} -
v_{\phi} \cos\theta) r \sin\theta e^{2i\phi} J_{22}^{[5]},
\\
\hspace{-0.6cm} F_{\theta}^{\rm GR} &\simeq& -i \kappa r^2
\sin\theta e^{2i\phi} J_{22}^{[6]} \nonumber
\\
\hspace{-0.6cm} & & +\, 3 \kappa ( -i v_{r} + v_{\phi} \sin\theta)
r \sin\theta e^{2i\phi} J_{22}^{[5]},
\\
\hspace{-0.6cm} F_{\phi}^{\rm GR} &\simeq& \kappa r^2 \sin\theta
\cos\theta e^{2i\phi} J_{22}^{[6]} \nonumber
\\
\hspace{-0.6cm} & & +\, 3 \kappa ( v_{r} \cos\theta - v_{\theta}
\sin\theta) r \sin\theta e^{2i\phi} J_{22}^{[5]}.
\end{eqnarray}
\end{subequations}

Now, let us apply the perturbative approach used in our paper to
check if our results match with Eq.~(\ref{ForceSphericJ22}). If we
linearize the multipole moments, $J_{lm}=
\hat{J}_{lm}+\delta^{(1)} \! J_{lm}$, where $\hat{J}_{lm}$
describes the multipole moment of the unperturbed star and
$\delta^{(1)} \! J_{lm}$ is the first-order Eulerian change in the
multipole moment, it is straightforward to show, using
Eqs.~(\ref{Jspherical}) and (\ref{UnperturbVeloc}), that
\begin{equation}
\hat{J}_{lm}=0\,;\qquad l=2; \, m=0,1,2,
 \label{unpertJ22:end}
\end{equation}
and the use of Eqs.~(\ref{Jspherical}) and
(\ref{r-mode:noGRforce}) yields
\begin{subequations}
 \label{deltaJ22:end}
\begin{eqnarray}
 \delta^{(1)} \! J_{22} &=& \alpha \Omega  \frac{\tilde{J}}{R}
\, e^{i\omega t} + {\cal O} (\alpha \Omega^3),
 \label{deltaJ22:enda} \\
\delta^{(1)} \! J_{21} &=& 0 + {\cal O} (\alpha \Omega^3),  \label{deltaJ22:endb}\\
\delta^{(1)} \! J_{22} &=& 0 + {\cal O} (\alpha \Omega^3),
\label{deltaJ22:endc}
\end{eqnarray}
\end{subequations}
together with
\begin{equation}
\delta^{(1)} \! J_{22}^{[n]} =(i \omega)^n  \delta^{(1)} \!
J_{22}.
 \label{deltaJ22:Derivative}
\end{equation}
So, the right-hand side of the first-order Eulerian change of
Eqs.~(\ref{relation:Sij-J2mb}) and (\ref{relation:Sij-J2mc}) is
zero in agreement with the assumptions (\ref{relation:Sij-J2mv2b})
and (\ref{relation:Sij-J2mv2c}) taken in
Refs.~\cite{LindTohVal2001,LindTohVal2002,GressmanEtAll2003}, and
the derivative of $\delta^{(1)} \! J_{22}$ satisfies
(\ref{deltaJ22:Derivative}) in agreement with the approximation
(\ref{J22:Derivative}) assumed in
Refs.~\cite{LindTohVal2001,LindTohVal2002,GressmanEtAll2003}.
Moreover, if we insert the value of $\delta^{(1)} \! J_{22}$ given
in Eq.~(\ref{deltaJ22:enda}) into the first-order Eulerian change
in the GR force we obtain an expression that agrees with our
result (\ref{deltaForce})\footnote{In fact, the value of $\kappa$
in Refs.~\cite{LindTohVal2001,LindTohVal2002,GressmanEtAll2003} is
two times ours (compare our Eq.~(\ref{Def:kappa}) with Eqs.~(A.8)
and (A.9) of Ref.~\cite{LindTohVal2002}). Their extra factor of 2
appears to be a typographical mistake. Indeed, with our value of
$\kappa$ we find an exact agreement with the instability timescale
of the $r$-mode in a spherical shell [see discussion after
Eq.~(\ref{ForceSpheric})] and we also get for the Newtonian star a
GR growth timescale that agrees with the one found in
Refs.~\cite{LindOwenMors1998, AndersKokkSchutz1999}. If we use the
value of $\kappa$ defined in
Refs.~\cite{LindTohVal2001,LindTohVal2002,GressmanEtAll2003} we
obtain a timescale that is twice the correct one in both
geometries.}.

\vfill



\begin{thebibliography}{99}
\bibitem{Chandra1970} S. Chandrasekhar, Phys. Rev. Lett. {\bf 24}, 611 (1970).
\bibitem{FriedSchutz1978} J. L. Friedman and B. F. Schutz, Astrophys.
J. {\bf 221}, 937 (1978); Astrophys. J. {\bf 222}, 281 (1978).
\bibitem{Lind1995} L. Lindblom, Astrophys. J. {\bf 438}, 265 (1995).
\bibitem{PapaPringle1978} J. Papaloizou and J. E. Pringle, Mon.
Not. R. Astron. Soc. {\bf 182}, 423 (1978).
\bibitem{ProvostBerthRocca1981} J. Provost, G. Berthomieu, and A. Rocca,
Astron. Astrophys. {\bf 94}, 126 (1981).
\bibitem{Saio1982} H. Saio, Astrophys. J. {\bf 256}, 717 (1982).
\bibitem{SmeyMar1983} P. Smeyers and L. Martens,
Astron. Astrophys.  {\bf 125}, 193 (1983).
\bibitem{Andersson1998} N. Andersson, Astrophys. J. {\bf 502}, 708 (1998).
\bibitem{FriedMors1998} J. L. Friedman and S. M. Morsink,
Astrophys. J. {\bf 502}, 714 (1998).
\bibitem{LindOwenMors1998} L. Lindblom, B. J. Owen, and  S. M. Morsink,
Phys. Rev. Lett. {\bf 80}, 4843 (1998).
\bibitem{Thorne1980} K. S. Thorne, Rev. Mod. Phys. {\bf 52} 299
(1980).
\bibitem{AndersKokkSchutz1999} N. Andersson, K. D. Kokkotas, and B. F. Schutz,
Astrophys. J. {\bf 510}, 846 (1999).
\bibitem{SpruitPhinney1998} H. Spruit and E. S. Phinney,
Nature {\bf 393}, 139 (1998).
\bibitem{Bildstein1998} L. Bildstein, Astrophys. J. {\bf 501}, L89
(1998).
\bibitem{AndersKokkSterg1999} N. Andersson, K. D. Kokkotas, and N. Stergioulas,
Astrophys. J. {\bf 516}, 307 (1999).
\bibitem{Chak2003} D. Chakrabarty et al, Nature {\bf 424}, 42
(2003).
\bibitem{OwenAl1998} B. J. Owen, L. Lindblom, C. Cutler, B. F.
Schutz, A. Vecchio, and N. Andersson, Phys. Rev. D{\bf 58}, 084020
(1998).
\bibitem{ArrasAl2003} P. Arras, E. E. Flanagan, S. M. Morsink, A.
K. Schenk, S. A. Teukolsky, and I. Wassermann, Astrophys. J. {\bf
591}, 1129 (2003).
\bibitem{RezzLambShap1} L. Rezzolla, F. K. Lamb, and S. L. Shapiro,
Astrophys. J. Lett. {\bf 531}, L141 (2000).
\bibitem{RezzLambShap2} L. Rezzolla, F. K. Lamb, D. Markovic, and
S. L. Shapiro, Phys. Rev. D {\bf 64}, 104013 (2001).
\bibitem{StergFont2001} N. Stergioulas and J. A. Font,
Phys. Rev. Lett. {\bf 86}, 1148 (2001).
\bibitem{LindTohVal2001} L. Lindblom, J. E. Tohline, and  M. Vallisneri,
Phys. Rev. Lett. {\bf 86}, 1152 (2001).
\bibitem{LindTohVal2002} L. Lindblom, J. E. Tohline, and M. Vallisneri,
Phys. Rev. D{\bf 65}, 084039 (2002).
\bibitem{GressmanEtAll2003} P. Gressman, L.-M. Lin, W.-M Suen,
N. Stergioulas, and J. L. Friedman, Phys. Rev. D {\bf 66}, 041303
(2002).
\bibitem{LevinUsh2001} Yu. Levin and G. Ushomirsky,
Mon. Not. R. Astron. Soc. {\bf 322}, 515 (2001).
\bibitem{Sa2004} P. M. S\'a, Phys. Rev. D {\bf 69}, 084001 (2004).
\bibitem{SaTome2004} P. M. S\'a and B. Tom\'e, Phys. Rev. D {\bf 71},
044007 (2005).
\bibitem{Blanchet1997} L. Blanchet, Phys. Rev. D{\bf 55}, 714 (1997).
\bibitem{Rezzolla1999} L. Rezzolla, M. Shibata, H. Asada,
T. W. Baumgarte, and S. L. Shapiro, Astrophys. J. {\bf 525}, 935
(1999).
\bibitem{Yoshida2000} S'i. Yoshida, S. Karino, S. Yoshida, and Y. Eriguchi,
Mon. Not. R. Astron. Soc. {\bf 316}, L1 (2000).
\bibitem{DiasSa} O. J. C. Dias and P. M. S\'a, in
preparation (2005).
\end{thebibliography}
\end{document}